\documentstyle[12pt]{article}
\newcommand {\da} {\downarrow}
\newcommand {\ua} {\uparrow}

\newcommand {\g} {\gamma}

\newcommand{\be}{\begin{equation}}
\newcommand{\ee}{\end{equation}}  

\newcommand{\kslash}{\mbox{$\not{\hspace{-0.8mm}K}$}}          
\newcommand{\qslash}{\mbox{$\not{\hspace{-0.8mm}q}$}}          
\newcommand{\pslash}{\mbox{$\not{\hspace{-0.8mm}P}$}}          
\newcommand{\vslash}{\mbox{$\not{\hspace{-0.8mm}v}$}}          
  
\begin{document}
\bibliographystyle{references}
\baselineskip 18pt
\pagenumbering{arabic}          
\begin{titlepage} 
\begin{flushright}
UTPT-98-03\\
MZ-TH/98-08
\end{flushright}

\begin{center} 

  {\large \bf Charmed Baryon Strong Coupling Constants in a Light-Front 
   Quark Model} 

 \vspace*{0.8cm}

 {\large
   Salam Tawfiq \hspace*{0.02cm}, \hspace*{0.04cm}  Patrick J. O'Donnell} 

\vspace*{0.4cm}

 {\large Department of Physics, University of Toronto \\
  60 St. George Street, Toronto Ontario, M5S 1A7, Canada }

  \vspace*{0.4cm}
  and \\
  \vspace*{0.4cm}

{\large J.G. K\"{o}rner}\\

  \vspace*{0.4cm}

{\large
  Institut f\"{u}r Physik, Johannes Gutenberg-Universit\"{a}t\\
  Staudinger Weg 7, D-55099 Mainz, Germany}\\
 \vspace*{0.5cm}
{\large February 1998}
       
 \vspace*{1cm}

\end{center} 
  
\begin {abstract}
  Light-Front quark model spin wave functions are employed to calculate 
  the three independent couplings $g_{\Sigma_c \Lambda_c \pi}$, 
  $f_{\Lambda_{c1} \Sigma_c \pi}$ and
    $f_{\Lambda^{*}_{c1} \Sigma_c \pi}$
  of S-wave to S-wave and P-wave to
  S-wave one-pion transitions. It is found that 
  $g_{\Sigma_c \Lambda_c \pi}=6.81\; {\rm GeV}^{-1}$, 
  $f_{\Lambda_{c1} \Sigma_c \pi}=1.16$ and 
  $f_{\Lambda^{*}_{c1} \Sigma_c \pi}=0.96\times 10^{-4}\; {\rm MeV}^{-2}$. 
  We also predict decay rates for specific strong transitions 
   of charmed baryons. 
 \end{abstract}
  
\end{titlepage}

\newpage

In the heavy quark  limit, the spin and parity of the heavy quark
and light  degrees of freedom  are  separately  conserved  in the
hadron.  In  addition,  strong  and  electromagnetic  transitions
among heavy  baryon  states are  transitions  solely of the light
quark system.  Therefore, heavy quark symmetry when  supplemented
by light flavour symmetries, such as $SU(2)$ or $SU(3)$ symmetry,
relate  these  decays. Explicit 
relations between the various decay couplings of heavy baryons were 
derived in the constituent quark model
 \cite{hks,py}. S-wave  to  S-wave  heavy  baryon  strong
decays,  for  instance,  are  determined  by  a  single  coupling
constant and two  independent  couplings are required to describe
single  pion  transitions  from  P-wave  to S-wave  states. 

  The coupling $g_{\Sigma_c \Lambda_c \pi}$ determines strong decays 
  among charmed baryon 
 ground states.
  Furthermore, single pion transitions from the first excited states into  
  ground state  
 are described in terms of two couplings $f_{\Lambda_{c1} 
 \Sigma_c \pi}$ 
 and $f_{\Lambda^{*}_{c1} \Sigma_c \pi}$. The
  $\Lambda_{c1}$ 
 and $\Lambda^{*}_{c1}$ represent
 the two excited states  
 discovered recently \cite{exp} with masses $2593 \; {\rm MeV}$ and 
 $2625 \; {\rm MeV}$ respectively.

 In a heavy baryon, a
   light diquark system with quantum numbers $j^P$ couples with a
   heavy quark with $J_Q^P=1/2^+$ to form a doublet with
   $J^P=(j\pm1/2)$. Heavy quark symmetry allows us to write down a general 
   form for the
   heavy baryon spin wave functions (s.w.f.) 
   \cite{hks,hklt} 
\begin{equation}\label{chigen}
\chi_{\alpha\beta\gamma}=(\phi_{\mu_1\cdots\mu_j})_{\alpha\beta}
\psi_{\gamma}^{\mu_{1} \dots\mu_{j}}(v)\;\;. 
\end{equation}
 Here, $v_{\mu}=\frac{P_{\mu}}{M} $ is the baryon four velocity, the spinor 
 indices \footnote{ We have ignored the isospin 
  indices which will be included in the transition amplitudes later on.} 
  $\alpha$ and $\beta$ refer to the
   light quark system and the index $\gamma$ refers to the heavy quark.
   The number of the Lorentz indices $\mu_j$ is determined by the light diquark
   system quantum number $j$ and is equal to $0$, $1$ and $2$ for S-wave 
   and P-wave baryon states.  
 In the heavy quark limit, the $\chi_{\alpha\beta\gamma}$ satisfy the 
   Bargmann-Wigner equation on the heavy quark index,
 \be
 (\vslash)^{\gamma^{\prime}}_{\gamma}\chi_{\alpha\beta\gamma^{\prime}}=
 \chi_{\alpha\beta\gamma} \;\; .
 \ee     
   The 
   light degrees of freedom spin wave functions
   $(\phi_{\mu_1\cdots\mu_j})_{\alpha\beta}(v)$
   are in general written in terms of the two bispinors 
   $[\chi^0]_{\alpha\beta}$ and 
   $[\chi^1_{\mu}]_{\alpha\beta}$. The matrix
   $[\chi^0]_{\alpha\beta}=[(\vslash +1)\gamma_5C]_{\alpha\beta}$, project out 
   a spin-0 object and is symmetric when interchanging $\alpha$ and 
   $\beta$. However, $[\chi^1_{\mu}]_{\alpha\beta}=[(\vslash +1)
    \gamma_{\perp \mu}C]_{\alpha\beta}$ which projects out a spin-1 object is 
    antisymmetric. Here, $C$ is the charge conjugation operator and 
      $\gamma^{\perp}_{\mu}=\gamma_{\mu}-\vslash v_{\mu}$.
 On the other hand
 the ``superfield" 
 $\psi_{\gamma}^{\mu_{1} \dots\mu_{j}}(v)$ stands for  
  the two spin wave
 functions corresponding to the two heavy quark symmetry degenerate 
 states with spin $j-1/2$ and $j+1/2$. They are
  generally written in terms of the Dirac spinor $u$ and 
  the Rarita-Schwinger spinor $u_{\mu}$.
 The S-wave heavy-baryon spin wave functions are 
 given by 
\be\label{Sstate}
(\phi^{\Lambda_Q})_{\alpha\beta}=(\chi^{0})_{\alpha\beta} \;\;\; {\rm ,} \;\;\;
(\psi^{\Lambda_Q})_{\gamma}=u_{\gamma} 
\ee
and
\be  
(\phi^{\mu \; \Sigma_Q})_{\alpha\beta}=({\chi}^{1,\mu})_{\alpha\beta}\;\;\; 
{\rm ,}\;\;\; (\psi^{\Sigma_Q}_{\mu})_{\gamma}=\left\{\begin{array}{r}
\frac{1}{\sqrt{3}}\gamma^{\perp}_{\mu} \gamma_{5}u \\
u_{\mu} \end{array} 
\right\}_{\gamma} \; .
\ee
For P-wave heavy baryon states, we shall use the relative momentum
  $K=\frac{1}{\sqrt{6}}(p_1+p_2-2p_3)$,  
 symmetric under interchang of the constituent light quark momenta
  $p_1$ and $p_2$, to represent the orbital excitation. The 
  $\Lambda_{Q1}$ degenerate state spin wave functions can be written as 
 \be \label{Pstate} 
(\phi^{\mu \; \Lambda_{Q1}})_{\alpha\beta}=({\chi}^{0}K^{\mu})_{\alpha\beta}\;\;\; 
{\rm ,}\;\;\; (\psi^{\Lambda_{Q1}}_{\mu})_{\gamma}=\left\{\begin{array}{r}
\frac{1}{\sqrt{3}}\gamma^{\perp}_{\mu} \gamma_{5}u \\
u_{\mu} \end{array} 
\right\}_{\gamma} \;.
\ee
  A more  
 detailed analysis with all heavy baryon P-wave spin wave functions 
 was presented in \cite{hks,hklt}. 

In the heavy quark  limit, we can write down the general form for
single pion transition amplitudes between heavy baryons
\begin{eqnarray}\label{mpi}
  M_{\pi}&=& \langle B_{Q^{'}}(P^{\prime}) \mid j_{\pi}(q) \mid B_{Q}(P) 
  \rangle
  \nonumber\\[2mm]
  &=&f_{B_{Q}B^{'}_{Q}\pi}
  {\bar \psi}^{ ' \nu_1 \dots \nu_{j_2}}(P^{\prime})
 (t^{\pi}(q))_{\nu_{1} \dots \nu_{j_{2}}}^{ \mu_1 \dots
       \mu_{j_1}}
   \psi_{\mu_{1} \dots
    \mu_{j_{1}}}(P) 
  \,,\label{trans}
  \end{eqnarray}
  with $j_{\pi}$ being the strong current, $f_{B_{Q}B^{'}_{Q}\pi}$ is the 
  appropriate strong coupling constant and the pion momentum 
  $q=P-P^{\prime}$.
  The light degrees of freedom transition tensors
  $(t^{\pi}(q))_{\nu_{1} \dots \nu_{j_{2}}}^{ \mu_1 \dots
       \mu_{j_1}}$
  of rank $(j_{1}+j_{2})$, built from 
  $g_{\perp\mu\nu}=g_{\mu\nu}-v_{\mu}v_{\nu}$ and the pion momentum, should 
  have the correct parity and
  project out
  the appropriate partial wave amplitude. 
  
  The $ \Sigma^{(*)}_c \rightarrow \Lambda_c\pi $, 
   $\Lambda_{c1} \rightarrow \Sigma_c\pi  $ and 
   $\Lambda^{*}_{c1} \rightarrow \Sigma_c\pi  $ covariant
   tensors $ (t^{\pi}(q))_{\nu_{1} \dots \nu_{j_{2}}}^{ \mu_1 \dots
       \mu_{j_1}}$ are $q_{\perp\mu}$, $g_{\perp \mu\nu}$ and 
  $q_{\perp\mu}q_{\perp\nu}$, with 
  $q_{\perp\mu}=q_{\mu}-v\cdot qv_{\mu}$, correspond to P-wave, S-wave 
  and D-wave transitions respectively. 
  Making use of the heavy baryon spin
  wave functions given in Eqs. (\ref{Sstate}-\ref{Pstate}) the strong 
  transition amplitudes,
   therefore,
   can be written as   
\begin{equation} \label{fp}
 \langle\Lambda (P^{\prime},{\lambda^{\prime}})
  | j_{\pi}(q) |\Sigma(P,{\lambda}) \rangle=\frac{1}{\sqrt{3}}
  g_{\Sigma_c\Lambda_c\pi} I_1
  \bar{u}(P^{\prime},{\lambda^{'}})\qslash_{\perp}\gamma_5 u(P,{\lambda}),
\end{equation} 
\be
 \langle\Lambda (P^{\prime},{\lambda^{\prime}})
  | j_{\pi}(q) |\Sigma^{*}(P,{\lambda}) \rangle=
  g_{\Sigma^{*}_c\Lambda_c\pi} I_1
  \bar{u}(P^{\prime},{\lambda^{'}})q_{\perp\mu}u^{\mu}(P,{\lambda}),
 \ee
 \begin{equation}\label{fs}  
 \langle\Sigma (P^{\prime},{\lambda^{\prime}})
 | j_{\pi}(q) |\Lambda_{c1}(P,{\lambda}) \rangle=
 f_{\Lambda_{c1}\Sigma_c\pi} I_3
 \bar{u}(P^{\prime},{\lambda^{'}})u(P,{\lambda}),
 \end{equation} 
 and 
 \begin{equation}\label{fd}
 \langle\Sigma(P^{\prime},{\lambda^{\prime}})
 | j_{\pi}(q) |\Lambda_{c1}^{*} (P,\lambda) \rangle=\frac{1}{\sqrt{3}}
 f_{\Lambda_{c1}^{*}\Sigma_c\pi} I_3
 \bar{u}(P^{\prime},{\lambda^{'}})\gamma_5\qslash_{\perp}u^{\mu}
 (P,{\lambda})q_{\perp \mu}\; ,
\end{equation}
 where
  $\lambda$ ($\lambda^{'}$)
     is the helicity of 
   the initial (final) spin-$\frac{1}{2}$ or spin-$\frac{3}{2}$ heavy 
   baryon. The  
   $I_1 \equiv I (6\rightarrow 3^{*}+\pi)$ and 
 $I_3 \equiv I(3^{*}\rightarrow 6+\pi)$
 are the appropriate group theoretical flavour factors.
 In fact, these are the only amplitudes 
 allowed by Lorentz invariance and parity conservation.  
  As was discussed in \cite{hks}, the S-wave coupling of Eq. (\ref{fs}) is 
  different from the one introduced in the HHCPT which is related to the scalar
  component of the axial vector current. 
 The matrix elements, Eqs. (\ref{fp}-\ref{fd}),
 can be transformed into their equivalent 
 effective chiral amplitudes \cite{py,xpt,cho,hdh} by replacing the pion
 momentum $q_{\mu}$ 
 by $-\partial_{\mu} \pi$ with
  the spinors 
 $u(p)$, $u_{\nu}(p)$ 
 and $\bar{u}(p)$ being replaced by the
 corresponding heavy baryon fields. The couplings 
  $g_{\Sigma_c\Lambda_c\pi}$ which is equal to $g_{\Sigma^{*}_c\Lambda_c\pi}$ 
  in the heavy quark limit, 
 $f_{\Lambda_{c1}\Sigma_c\pi}$ and $ f_{\Lambda^{*}_{c1}\Sigma_c\pi}$ are 
 related respectively to $g_2$, $h_2$ and $h_8$ defined in  
  the Heavy Hadron Chiral Perturbation Theory (HHCPT) \cite{py,xpt,cf} such 
  that $ g_{\Sigma_c\Lambda_c\pi}=\frac{\sqrt{3}g_2}{\sqrt{2}f_{\pi}}$, 
 $ f_{\Lambda_{c1}\Sigma_c\pi}=\frac{\sqrt{2}h_2}{f_{\pi}}E_{\pi}$ and
  $ f_{\Lambda^{*}_{c1}\Sigma_c\pi}=\frac{6 h_8}{ \sqrt{5}f_{\pi}}$ with 
 $f_{\pi}=0.093 \; {\rm GeV}$. 

 The single-pion decay rates are 
  calculated using the general formula   
  \be\label{rate}
  \Gamma = \frac{1}{2J_{1}+1} \quad \frac{ \mid \vec{q} \mid}{8 \pi
  M_{B_Q}^{2}}\sum_{spins} \mid M^{\pi} \mid^{2},
  \ee
  with $\mid \vec{q} \mid$ being the pion momentum in the rest frame 
  of the decaying baryon. Using Eqs. (\ref{fp}-\ref{fd})
 and (\ref{rate}), we get 
\be\label{prate}
\Gamma\left( \Sigma_c \rightarrow \Lambda_c \pi \right)=
 \Gamma\left( \Sigma^{*}_c \rightarrow \Lambda_c \pi \right)
  = g^2_{\Sigma^{(*)}_c\Lambda\pi}I_1^2
  \frac{{\mid \vec{q} \mid}^3}{6\pi} \frac{M_{\Lambda_c}}{M_{\Sigma^{(*)}_c}} 
\ee
\be\label{srate}
\Gamma\left(\Lambda_{c1}  \rightarrow  \Sigma_c \pi  \right) 
     = f^2_{\Lambda_{c1}\Sigma\pi} I_3^2  
	 \frac{\mid \vec{q} \mid}{4\pi} \frac{M_{\Sigma_c}}{M_{\Lambda_{c1} }}
\ee
\be\label{drate}
\Gamma\left(\Lambda^{*}_{c1}  \rightarrow  \Sigma_c \pi \right)
   = f^2_{\Lambda^{*}_{c1}\Sigma\pi} I_3^2
    \frac{{\mid \vec{q} \mid}^5}{36\pi} \frac{M_{\Sigma_c}}
    {M_{\Lambda^{*}_{c1} }}
\ee
Assuming that the 
  width of 
 $\Sigma_c$, $ \Lambda_{c1} $ and 
 $ \Lambda^{*}_{c1} $ are saturated by strong decay channels one can 
 estimate the values 
 of the three couplings using the experimental decay rates.
 Taking $\Gamma_{ \Sigma_c^{* ++}\rightarrow \Lambda_c^{+} \pi^{+}}=
  17.9^{+3.8}_{-3.2} \; {\rm MeV}$, 
  $\Gamma_{ \Sigma_c^{*0}\rightarrow \Lambda_c^{+} \pi^{-}}=
  13.0^{+3.7}_{-3.0}\; {\rm MeV}$
 reported by CLEO \cite{exp} Eq. (\ref{prate})
 can be used to determine 
 \footnote{ Numerical values for the masses will be
  taken from Table 1 of \cite{cf}. In this analysis,
  which is similar to those done in \cite{py,xpt,cf}, we use the 
  updated data reported in the Review of Particle Physics \cite{PDG}.} 
  the coupling $g_{\Sigma_c\Lambda_c\pi}$. One, therefore, respectively gets 
\be\label{g-value}
  g_{\Sigma_c\Lambda_c\pi}=8.03^{+1.97}_{-1.92} \; {\rm GeV}^{-1} \;
\ee
 and
\be
g_{\Sigma_c\Lambda_c\pi}=6.97^{+1.84}_{-1.74}\; {\rm GeV}^{-1} \;
\ee
  These values, in return, give the analogous HHCPT coupling 
  $g_2=0.61^{+0.15}_{-0.14}$ and 
  $g_2=0.53^{+0.14}_{-0.13}$ defined in \cite{py,xpt}.
   
   To estimate $f_{\Lambda_{c1}\Sigma\pi}$ we use the
    Particle Data Group \cite{PDG} average
   value for 
   $ \Lambda_{c_1}(2593)$ width 
   which is 
   $\Gamma_{\Lambda_{c_1}(2593)}=3.6^{+2.0}_{-1.3}\; {\rm MeV} $ and 
   Eq. (\ref{srate}) 
   to obtain  
 \be 
  f_{\Lambda_{c1}\Sigma\pi}=1.11^{+0.31}_{-0.20}.
 \ee
  The corresponding HHCPT coupling constant $h_2$ is calculated to be 
  $h_2=0.73^{+0.20}_{-0.13}$.

  Finally, taking the upper bound on the $\Lambda^{+}_{c_1}(2625)$ width 
  obtained by CLEO \cite{exp} ($\Gamma_{\Lambda^{+}_{c_1}(2625)}
  < 1.9\; {\rm MeV}$), Eq. (\ref{drate}) gives 
  \be \label{f-value}
  f_{\Lambda^{*}_{c1}\Sigma\pi}=1.66\times 10^{-4}\; {\rm MeV}^{-2} \;.
  \ee
  The value of the HHCPT D-wave coupling $ h_8 $ is determined to be 
  $h_8=5.75\times 10^{-3}\; {\rm MeV}^{-1} $.  
  The uncertainty in the values of the couplings is dominated 
  by the experimental errors 
  in the decay rates and in the baryons masses. 

 Theoretically, to calculate the three couplings one needs to evaluate the 
 matrix elements of $j_{\pi}(q)$ 
 in Eqs. (\ref{fp}), 
 (\ref{fs}) and (\ref{fd}) at $\stackrel{\rightarrow}{{\bf q}}^2=0$ in 
 an appropriate frame of reference. 
 The Light-Front (LF) formalism \cite{lc} provides a consistent relativistic 
 theory for composite systems with a fixed number of constituent. 
 The other essential fact is that the Melosh 
 rotation \cite{melosh} is
 already included in the LF 
 spinors which is important when calculating form  
 factors. Therefore,
 we shall employ (LF) wave 
 functions to describe the initial and final heavy baryons.  
 
 Without loss of
 generality, 
  we choose to work in a 
  Drell-Yan frame where the initial baryon momentum 
  $P^{\mu}=\left(P^+,\frac{M^2}{P^+}, 
  {\bf 0}_{\perp} \right)$ and 
  the pion momentum 
  $q^{\mu}=\left(0,\frac{M^2-M^{' 2}-{\bf q}_{\perp}^2 }{P^+},
    {\bf q}_{\perp} \right)$.
  With the aid of the
 Light Front spinors and matrix elements of the appropriate gamma matrices 
  defined in the appendix,
   which become even simpler since more elements will vanish in
   this frame, the three independent couplings are given by 
 \begin{equation}\label{g2}
 g_{\Sigma_c \Lambda_c \pi}=
 -\frac{2\sqrt{3M_{\Lambda_c}M_{\Sigma_c}}}{(M_{\Sigma_c}^2-M_{\Lambda_c}^2)}
 \langle \Lambda (P^{\prime},{\ua})
|{\hat j}_{\pi}(0)|\Sigma(P,{\ua}) \rangle
\end{equation}
 \begin{equation}\label{h2}
 f_{\Lambda_{c1} \Sigma_c \pi}=
 \langle\Sigma (P^{\prime},{\ua})
 | {\hat j}_{\pi}(0)|\Lambda_{c1}(P,{\ua}) \rangle
 \; ,
 \end{equation}
 and
 \be\label{h8}
 f_{\Lambda^{*}_{c1} \Sigma_c \pi}=
 \frac{3\sqrt{2}}{ (M_{\Lambda^{*}_{c1}}-M_{\Sigma})^2}
 \frac{M_{\Lambda^{*}_{c1}}^2}{ (M_{\Lambda^{*}_{c1}}^2-M_{\Sigma_c}^2)} 
\langle \Sigma (P^{\prime},{\ua})
 |{\hat j}_{\pi}(0)|\Lambda^{* }_{c1}(P,\frac{1}{2}) \rangle
 \ee
   The LF matrix elements of the strong transition current ${\hat j}_{\pi}(q)$  
   between 
   heavy baryon states are
  given by
 \[
\langle B^{\prime}(P^{\prime},\lambda^{\prime})|{\hat j}_{\pi}(q) |
 B(P,{\lambda})\rangle=
\int [dx_i] [d^2{\bf p}_{\perp i}]\sum_{\lambda_i,\lambda_i^{\prime}}
 \psi_{B^{\prime}}^{\dagger }(x_i^{\prime},{\bf p}^{\prime}_{\perp i},
 \lambda_i^{\prime};\lambda^{\prime}) \]
 \be
(\sum_{j=1,2}{\bar u}(p_j^{\prime},{\lambda_j^{\prime}}) \hat{j}_{\pi}(q)
u(p_j,{\lambda_j}))
 \psi_{B}(x_i,{\bf p}_{\perp i},\lambda_i;{\lambda}),
\ee
 where 
 $\psi_{B}(x_i,{\bf p}_{\perp i},\lambda_i;{\lambda})$ and
 $\psi_{B^{\prime}}^{\dagger }(x_i^{\prime},
 {\bf p}^{\prime}_{\perp i},\lambda_i^{\prime};\lambda^{\prime})$ 
 are the initial and final heavy baryon wave functions explicitly given in 
 Eq. (\ref{psitot}) below.  
 In the constituent quark model the pion
 is assumed to be emitted by each of the light quarks and the heavy quark 
 is not affected.
  Therefore,
 the strong current is
 resolved into constituent quark transitions and its appropriate
 operator $\hat{j}_{\pi}(q)$ can be written as
 \be\label{j-p}
 \left(\hat{j_{\pi}}\right)^{\alpha\beta}_{\alpha^{'}\beta^{'}} =\frac{1}{2}
 \left((\gamma_{5})^{\alpha}_{\alpha^{'}}
 \delta^{\beta}_{\beta^{'}} 
  +\delta^{\alpha}_{\alpha^{'}}(\gamma_5)
  ^{\beta}_{\beta^{'}}
  \right)
 \ee
 The most difficult
  part in calculating these form factors, however, is related to the
  choice of the form of initial
  and final baryon wave functions. One of the advantages of Light-Front (LF) 
  formalism
  \cite{lc} is that,  
 all Fock-state wave functions $\Psi(x_i,p_{\perp i},\lambda_i;{\lambda})$, 
 with helicity $\lambda$ and constituent transverse momenta 
 $p_{\perp i}$, tend 
 to vanish when the LF energy $\epsilon$ becomes infinitely
 large. 
 This feature, is very much similar to the so called
 "valence"  constituent quark model where the dynamics are dominated by the
 valence quark structure.

  In the LF formalism the total baryon spin-momentum distribution function 
  can be written in  
  the following general form  
\be\label{psitot}
\Psi(x_i,{\bf p}_{\perp i},\lambda_i;{\lambda})=
\chi(x_i,{\bf p}_{\perp i},\lambda_i;{\lambda}) \psi(x_i,{\bf p}_{\perp i}).
 \ee
 Here, $\chi(x_i,{\bf p}_{\perp i},\lambda_i;{\lambda})$ and 
 $\psi(x_i,{\bf p}_{\perp i})$ represent the spin and momentum distribution 
 functions respectively and the longitudinal-momentum fraction 
\be
 x_i=\frac{p_i^+}{P^+} \;\; {\rm with} \;\; \sum_{i=1}^{3}x_i=1.
\ee
  These functions are normalized such that
 \be
 \int [dx_i] [d^2{\bf p}_{\perp i}]\sum_{\lambda_i}
 \psi_{B^{'}}^{\lambda^{'}\dagger }(x_i,{\bf p}_{\perp i};\lambda_i)
 \psi_{B}^{\lambda}(x_i,{\bf p}_{\perp i};\lambda_i)=\delta_{\lambda
 \lambda^{\prime}},
  \ee
  with
  \begin{equation}
 [dx_i]=\prod_i dx_i \delta(1-\sum_i x_i)\;, \;[d^2{\bf p}_{\perp i}]=\prod_i
 d^2{\bf p}_{\perp i}16\pi^3 \delta^2(\sum_i {\bf p}_{\perp i})
 \end{equation}
 Assuming factorization of the longitudinal $\phi(x_i)$ and
   transverse momentum distribution functions,  
   $\psi(x_i,{\bf p}_{\perp i})$ 
 can be written as 
 \be\label{mom} 
 \psi(x_i,{\bf p}_{\perp i})=\phi(x_i)exp\left[-\frac{ \stackrel{\rightarrow}{ 
{\bf k}}
 ^2}
 {2\alpha_{\rho}^2}
  -\frac{ \stackrel{\rightarrow}{{\bf K}}^2}{2\alpha^2_{\lambda}}
  \right]\; .
\ee
  The transverse component of the momentum distribution are assumed to be 
  harmonic oscillator eigenfunctions
 with $\alpha_{\rho}$ and $\alpha_{\lambda}$ controlling
 the confinement of quarks in the heavy baryon.
 The momenta $\stackrel{\rightarrow}{\bf k}$ and
  $\stackrel{\rightarrow}{\bf K}$, corresponding to the 
  nonrelativistic
  three body momenta ${\bf k}_{\rho}$ and ${\bf k}_{\lambda}$, 
  are the transverse component of the covariant vectors 
\be
  k=\frac{1}{\sqrt{2}}(
  p_1-p_2) \;\; , \;\;
  K=\frac{1}{\sqrt{6}}
 ( p_1+ p_2
 -2 p_3). 
 \ee
 These harmonic oscillator functions were used 
 successfully in 
 \cite{cik} to predict masses and decay rates of ground state and excited 
 charmed baryons. They were also employed to calculate baryon magnetic 
 moments \cite{ck} and to calculate the Isgur-Wise function for 
 $\Lambda_Q$ semileptonic decay \cite{scora} in a relativistic 
  quark model. The choice of the relative momenta 
 $k$ and $K$ are also 
 convenient
 for keeping 
 track of the exchange 
 symmetry for the light degrees of freedom spin wave functions. They will 
 be used later on to write down 
 an explicit form for heavy baryon P-wave spin functions.   
 
   In the heavy quark limit, the heavy baryon longitudinal momentum 
   distribution functions $\phi(x_i)$ 
  are expected to   
  have most of their strength in the
  neighborhood of the mean values $\bar{x}_Q=\frac{m_Q}{M}$. In the weak
  binding \cite{hkt} or valence approximation \cite{vqm} the longitudinal 
  velocity of the 
  constituent quarks are the same. One therefore expects that  
 also for the light quarks the distribution is peaked fairly sharply around the
 equal velocity point $\bar{x}_i=\frac{m_i}{M}$ with $i=1$ and $2$. 
 Therefore, we can
 assume 
 \be\label{long}
 \phi(x_i)=\prod_{i=1}^{3} \delta(x_i-\bar{x}_i)
 \ee

 In the equal velocity assumption \cite{hkt,vqm}
 one may use the two projection operators
 $[\chi^0]_{\alpha\beta}$ and 
 $[\chi^{1,\mu}]_{\alpha\beta}$, defined earlier,
to 
write 
down the    
 spin-dependent functions. 
   The $\Lambda_Q$-like baryons spin wave function
   $\chi_{\Lambda_Q}(x_i,{\bf p}_{\perp i},\lambda_i;\lambda)$
 must be antisymmetric when interchanging the light quark indices and
 is given by 
 \be 
 \chi_{\Lambda_Q}(x_i,{\bf p}_{\perp i},\lambda_i;{\lambda})=
 \bar{u}^{\alpha_1}(p_1,{\lambda_1})\bar{u}^{\alpha_2}
 (p_2,{\lambda_2}) \bar{u}^{\alpha_3} (p_3,{\lambda_3})
 [\chi^0]_{\alpha_1\alpha_2} u_{\alpha_3}(P,{\lambda}), \nonumber
 \ee
 here, the LF spinors $u^{\alpha_i}(p_i,{\lambda_i})$ describe the  
 constituent quarks with momentum $p_i$ and helicity $\lambda_i$ and 
 $u^{\alpha}(P,{\lambda})$ refers to the $\Lambda_Q$-like baryon with 
 momentum $P$ and helicity $\lambda$. 
 $\chi_{\Lambda_Q}(x_i,{\bf p}_{\perp i},\lambda_i;\lambda)$ can be rewritten 
 in a more convenient form
 \be\label{chilam}
 \chi_{\Lambda_Q}(x_i,{\bf p}_{\perp i},\lambda_i;\lambda)=\bar{u}
( p_{1},{\lambda_1})[(\pslash +M_{\Lambda})\gamma_5]\nu
( p_{2},{\lambda_2})
 \bar{u}(p_3,{\lambda_3})u(P,{\lambda}).
  \ee
 For the $\Sigma_Q$-like baryon, the spin wave functions are symmetric in the 
 light quark indices and have the form 
 \be\label{chisig2}
 \chi_{\Sigma_Q}(x_i,{\bf p}_{\perp i},\lambda_i;{\lambda})=\bar{u}
 (p_1,{\lambda_1})[(\pslash +M_{\Lambda})\gamma_{\perp}^{\mu}]\nu
 (p_2,{\lambda_2})
 \bar{u}(p_3,{\lambda_3})\gamma_{\perp \mu}\gamma_5 
 u(P,{\lambda}),
 \ee
  The two relative momenta $k$ and 
$K$ can be used to specify the spin wave functions for heavy baryon 
resonances. The excited states $\Lambda_{Q1}$, with $J^P=\frac{1}{2}^-$, and 
 $\Lambda^{*}_{Q1}$, with $J^P=\frac{3}{2}^-$,  
  have spin functions of
the following forms 
\be\label{chilam12}
 \chi_{\Lambda_{QK1}}(x_i,{\bf p}_{\perp i},\lambda_i;{\lambda})=
 \bar{u}
(p_1,{\lambda_1})[(\pslash +M_{\Lambda_{c1}})\gamma_5]\nu(p_2,{\lambda_2})
 \bar{u}(p_3,{\lambda_3})\kslash  \gamma_5 
 u(P,{\lambda}),
\ee
and 
\be\label{chilam32}
\chi_{\Lambda^{*}_{QK1}}(x_i,{\bf p}_{\perp i},\lambda_i;{\lambda})=
 \bar{u}
 (p_1,{\lambda_1})[(\pslash +M_{\Lambda^{*}_{c1}})\gamma_5]\nu
 (p_2,{\lambda_2})
 \bar{u}(p_3,{\lambda_3}) K_{\mu}  
 u^{\mu}(P,\lambda) \;\; .
\ee
  One can obtain the spin wave functions for the 
 corresponding antisymmetric excited states by replacing $K$ with $k$. Explicit 
 forms for the spinors $u(p,\lambda)$ and $u^{\mu}(p,\lambda)$ and anti 
 spinors $\nu(p,\lambda)$ in 
 the LF formalism are given in the appendix. 
 
 Since there are two free parameters in our model, namely, the oscillator 
 couplings 
 $\alpha_{\rho}$ and $\alpha_{\lambda}$ one, therefore, expects that the 
  predictions made
 will depend mainly on these two parameters. The numerical values for 
 the constituent quark masses are taken to be $m_u=m_d=0.33 {\rm \; GeV}$, 
 $m_c=1.51 {\rm \; GeV}$ and those for
  $\alpha_{\rho}$ and $\alpha_{\lambda}$ are 
  $\alpha_{\rho}=0.40\; {\rm GeV}/c$ and $\alpha_{\lambda}=0.52 
 \; {\rm GeV}/c$.
  The same values for the oscillator 
 couplings were chosen to fit the $\Lambda$ baryon masses \cite{scora}. 
 However, one would expect that these values might slightly change for the 
 $\Xi$ 
 baryons since the constituent quarks are not the same as those in 
 the $\Lambda$ 
 and $\Sigma$ baryons. We shall postpone the study of the effect of these 
 parameters for a future work since the sensitivity of the 
 decay rates to the $\alpha$ values 
 is such that a $10\%$ increase results in about $(5-8)\%$ change in the 
 calculated 
 decay rates.
 
 To evaluate the integrals in Eqs. (\ref{g2}-\ref{h2}) we 
 introduce the relative momentum variables
\be
{\bf \zeta}_{\perp}=\frac{x_2 {\bf p}_{\perp 1}-x_1{\bf p}_{\perp 2}}{x_1+x_2}
\;\; ,
 \; \;{\bf \eta}_{\perp}=(x_1+x_2){\bf p}_{\perp 3}-x_3({\bf p}_{\perp 1}+
{\bf p}_{\perp 2})
 \ee
 These variables have the crucial property of being space-like four-vectors
  because of the vanishing of the invariant
  + component ($\zeta^+=\eta^+=0$). The momentum conservation relations are
\be
 x_i M=x_i^{\prime} M^{\prime}
\ee
and if the pion is emitted by quark number 1, we also have 
\be
{\bf \zeta}_{\perp}^{\prime}={\bf \zeta}_{\perp}-\frac{x_1}{x_1+x_2}{\bf 
q}_{\perp} \; \; {\rm and} \; \;
{\bf }\eta_{\perp}^{\prime}=
{\bf \eta}_{\perp}-x_3{\bf q}_{\perp}.
\ee
 Using Eqs.(\ref{mom}), (\ref{long}) and (\ref{chilam}-\ref{chilam32})
 the three charmed baryons strong couplings 
 $g_{\Sigma_c\Lambda_c\pi}$, $f_{\Lambda_{c1}\Sigma_c\pi}$ and 
 $f_{\Lambda^{*}_{c1}\Sigma_c\pi}$ are calculated to be 
 \be\label{theory}
 g_{\Sigma_c\Lambda_c\pi}=6.81\; {\rm GeV}^{-1} \; \; , \;\; 
   f_{\Lambda_{c1}\Sigma_c\pi}=1.16 \;\; ,\;\;
 f_{\Lambda^{*}_{c1}\Sigma_c\pi}=0.96\times 10^{-4}\; {\rm MeV}^{-2} \;. 
\ee
 These are in nice agreement with our 
 earlier fit to the upgraded CLEO measurements for 
 $\Gamma_{\Sigma_c^*\rightarrow\Lambda_c}$, 
 $\Gamma_{\Lambda_{c_1}(2593)\rightarrow \Sigma_c}$ and 
 $\Gamma_{\Lambda^{*}_{c_1}(2593)\rightarrow \Sigma_c}$
 strong decay rates. 
 The corresponding HHCPT couplings are determined using the values in
 Eq. (\ref{theory}); 
\be
 g_2=0.52  \; \; {\rm ,} \;\;
 h_2=0.54 \;\; {\rm ,} \;\; h_8=3.33\times 10^{-3}{\rm MeV}^{-1}\; .
 \ee 

 Having the three independent couplings in hand, we are now in a position to 
 predict charmed baryons strong decay rates. Ground state transitions are 
 saturated by P-wave transitions which can be  
  calculated using the value of $g_{\Sigma_c\Lambda_c\pi}$
  and Eq. (\ref{prate}). On the other hand, transitions from the first excited
  states are S-wave or D-wave transitions and their decay rates  
  are predicted using 
  Eq. (\ref{srate}) and Eq. (\ref{drate}) respectively. 
   These decay rates are summarized in 
   Table \ref{tab1} as well as the experimental values presented in 
   the updated version of the Review of Particle Physics \cite{PDG}.     
 
 From Table \ref{tab1}, one notes that the strong width of 
 $\Sigma^{*}_c$ is about seven to eight times larger than
 the width of its spin-$\frac{1}{2}$ partner $\Sigma_c$. These values are 
 within the range of the CLEO measurements. The 
 $\Xi^{* 0}_c$ and $ \Xi^{* +}_c $ strong decay width 
  are within the
  current upper bound obtained by CLEO.

 The $\Lambda_{c1}(2593)$ decay width receives contributions from both its 
 single-pion decay to $\Sigma_c$ and from
 decaying to 
 $\Lambda_c$ via a two-pion transition. The two-pion contribution was 
 computed in \cite{cho,hdh} 
 with the result $\Gamma_{\Lambda_{c1}(2593) \rightarrow \Lambda_c \pi\pi}=
 2.5 \; {\rm MeV}$. Hence, the total decay rate is 
 $ \Gamma_{\Lambda_{c1}(2593)}=6.49\; 
 {\rm MeV}$ which 
 is still consistent with the CLEO result $\Gamma_{\Lambda_{c1}(2593)}=
 3.6^{+2.0}_{-1.3}\; {\rm MeV}$. Actually, there is also a negligible 
 D-wave single-pion 
 contribution to the $\Lambda_{c1}(2593)$ width.  
 
 We also predict the S-wave branching ratios of 
 $\Xi_{c1}(2815) \rightarrow \Xi_c^{* 0} \pi^{+} $ 
 to $\Xi_{c1}(2815) \rightarrow \Xi_c^{* +} \pi^{0}$ 
 to be $67 \%$ and $33 \%$ respectively. The S-wave $\Xi_{c1}(2815)$ decay 
 width receives an extra $2\%$ contribution from D-wave modes
  giving a total width  
  $\Gamma_{\Xi_{c1}(2815)}=
 7.67 \; {\rm MeV}$. This value is about three times higher than the upper
 bound obtained 
 by CLEO $\Gamma_{\Xi_{c1}(2815)}<
  2.4 \; {\rm MeV}$. 
 
 Finally, the strong decay width of $\Lambda^{*}_{c1}(2625)$, the 
 spin-$\frac{3}{2}$ partner of 
 $ \Lambda_{c1}(2593)$, is 
 saturated by D-wave transitions to $\Sigma_c$ and by two-pion decay to 
 $\Lambda_c$. 
 Adding the contribution from two-pion decay 
 $\Gamma_{\Lambda^{*}_{c1}(2625) \rightarrow \Lambda_c \pi\pi}=0.035 \; {\rm MeV}$,
 calculated in \cite{hdh},
 one gets $\Gamma_{\Lambda^{*}_{c1}(2625)}=2.19 \; {\rm MeV}$ 
 which is close to the upper limit obtained by CLEO 
 $\Gamma_{\Lambda^{*}_{c1}(2625)}<1.9 \; {\rm MeV}$. 

 To summarize, we constructed Light-Front (LF) quark model functions 
 with a factorized harmonic oscillator transverse momentum component and 
 a longitudinal component given by Dirac delta-functions.
 The spin wave function are the LF generalization of the conventional 
 constituent quark model spin-isospin functions. These bound state distribution 
 functions were used to calculate the strong couplings for 
 $\Sigma_c\rightarrow\Lambda_c\pi$,
 $\Lambda_{c1}\rightarrow\Sigma_c\pi$ and 
 $\Lambda^{*}_{c1}\rightarrow\Sigma_c\pi$ decay modes which correspond           
 to P-wave, S-wave and D-wave transitions respectively.  
 The LF quark model predictions for the numerical 
 values of these couplings Eq. (\ref{theory}) are in good 
 agreement with estimates obtained using the 
 available experimental data Eqs. (\ref{g-value}-\ref{f-value}).
 Like other models, our results will mainly depend on the choice of the 
 free parameters which are the harmonic oscillator constants 
 $\alpha_{\rho}$ and $\alpha_{\lambda}$. The decay rates are also sensitive to 
 the numerical values of the masses of the heavy baryon states and some of 
 these masses have not been measured with high accuracy. We hope in the near 
 future our results will be confirmed by the new experimental data.    
 \section*{Acknowledgments}
   One of us S. T. would like to thank 
  Patrick J. O'Donnell and the Department of Physics, University of 
  Toronto for hospitality. This 
  research was supported in part by the National Sciences and Engineering 
  Research Council of Canada.
\begin{table}
\caption{\label{tab1} Decay rates for charmed baryon states.}
 \vspace{5mm}
 \renewcommand{\baselinestretch}{1.2}
 \small \normalsize
 \begin{center}
 \begin{tabular}{|c|c|c|}
 \hline \hline
 $B_Q\rightarrow B^{\prime}_{Q}\pi$ & $\Gamma \; ({\rm MeV}) $ & 
 $ \Gamma_{expt.} \; ({\rm MeV}) $  \\
 \hline\hline 
\multicolumn{3}{|l|}{P-wave transitions} \\ 
\hline  
$\Sigma^{+}_{c} \rightarrow \Lambda_c\pi^{0} $ &  $ 1.70  $ & $  $   \\ 
$\Sigma^{0}_{c} \rightarrow \Lambda_c\pi^{-} $ &  $ 1.57   $ & $  $   \\ 
$\Sigma^{++}_{c} \rightarrow \Lambda_c\pi^{+} $ &  $ 1.64 $ & $  $   \\ 
\hline 
 $\Sigma^{*0}_{c} \rightarrow \Lambda_c\pi^{-} $ &  $ 12.40 $ &  
 $ 13.0^{+3.7}_{-3.0} $   \\     
 $\Sigma^{* ++}_{c} \rightarrow \Lambda_c\pi^{+} $ &  $ 12.84 $ &  
 $ 17.9^{+3.8}_{-3.2} $   \\  
\hline
$\Xi^{*0}_{c} \rightarrow \Xi^{0}_c\pi^{0} $ &  $  0.72 $ & $ <5.5 $   \\
$\Xi^{*0}_{c} \rightarrow \Xi^{+}_c\pi^{-} $ &  $  1.16 $ &   \\
\hline
$\Xi^{*+}_{c} \rightarrow \Xi^{0}_c\pi^{+} $ &  $ 1.12  $ & $ <3.1 $    \\
$\Xi^{*+}_{c} \rightarrow \Xi^{+}_c\pi^{0} $ &  $ 0.69 $ & $  $   \\
\hline \hline 
\multicolumn{3}{|l|}{S-wave transitions} \\ 
\hline  
$\Lambda_{c1}(2593) \rightarrow \Sigma^{0}_c\pi^{+} $ & $2.61  $ & $ $  \\
$\Lambda_{c1}(2593) \rightarrow \Sigma^{+}_c\pi^{0} $ & $1.73  $ & 
$ 3.6^{+2.0}_{-1.3}$  \\
$\Lambda_{c1}(2593) \rightarrow \Sigma^{++}_c\pi^{-} $ & $2.15  $ & $ $  \\
 \hline 
$\Xi^{*}_{c1}(2815) \rightarrow \Xi^{*0}_c\pi^{+} $ &  $ 4.84  $ & 
$ \Gamma_{\Xi^{*}_{c1}}<2.4 $   \\
$\Xi^{*}_{c1}(2815) \rightarrow \Xi^{*+}_c\pi^{0} $ &  $ 2.38  $ & $  $   \\
\hline \hline 
\multicolumn{3}{|l|}{D-wave transitions} \\ 
\hline  
$\Lambda^{*}_{c1}(2625) \rightarrow \Sigma^{0}_c\pi^{+} $ & $ 0.77 $ & $  $ \\
$\Lambda^{*}_{c1}(2625) \rightarrow \Sigma^{+}_c\pi^{0} $ & $ 0.69 $ & 
$\Gamma_{\Lambda^{*}_{c1}}<1.9 $  \\
$\Lambda^{*}_{c1}(2625) \rightarrow \Sigma^{++}_c\pi^{-} $ & $ 0.73 $ & $ $  \\
\hline 
$\Xi^{*}_{c1}(2815) \rightarrow \Xi^{0}_c\pi^{+} $ & $ 0.30 $ & $ $  \\
$\Xi^{*}_{c1}(2815) \rightarrow \Xi^{+}_c\pi^{0} $ & $ 0.15 $ & $ $  \\
\hline \hline  
\end{tabular}
 \renewcommand{\baselinestretch}{1}
 \small \normalsize
 \end{center}
 \end{table}             
 
\newpage
\setcounter{equation}{0}
\def\theequation{A.\arabic{equation}}
\section*{ Appendix }
In this appendix explicit forms for Dirac spinors $u(p,\lambda)$ and  
Rarita-Schwinger
spinors $u^{\mu}(p,\lambda)$ in the Light-Front (LF) formalism are 
presented. Previously, the spin-$\frac{3}{2}$ wave functions have only been 
given in the canonical form 
\cite{fg}. We shall, also, give 
matrix elements of some useful 
$\gamma$-matrices between LF spinors.
The standard representation of $\gamma$ matrices is used. 
\begin{equation}
\g_{0}=\left[\matrix{I&0\cr 0&-I\cr}\right],\;\;\;
\g_{i}=\left[\matrix{0&\sigma^i\cr -\sigma^i&0\cr}\right],\;\;\;
\g_{5}=\left[\matrix{0&I\cr I&0\cr}\right] 
\end{equation}
where $\sigma^i$ being the usual Pauli matrices.

The spin-$\frac{1}{2}$ LF spinors $u_{\lambda} ( p )$  
 with four momentum $p=(p^+,p^-,{\bf p_{\perp} })$ and helicity    
 $\lambda=(\ua or \da)$ are given by \cite{vqm,cc} 
\begin{equation}
u(p,{\ua}) = \frac{1}{2\sqrt{mp^+}} 
\left[\matrix{p^+ + m\cr p^r \cr p^+ -m \cr p^r \cr}\right], \;\;\;
u(p,{\da}) = \frac{1}{2\sqrt{mp^+}} 
\left[\matrix{-p^l  \cr p^+ +m \cr p^l \cr -(p^+ -m) \cr }\right],
\end{equation}
here, we have defined $p^{l} = p_x - i p_y$ and 
$p^{r} = p_x + i p_y$. Similarly,  
 the anti spinors $\nu_{\lambda}(p)$ have the form 
\begin{equation}
\nu(p,\uparrow) = \frac{1}{2\sqrt{mp^+}} 
\left[\matrix{-p^l \cr p^+ -m \cr p^l \cr -(p^+ +m)\cr }\right], \;\;\;
\nu(p,\downarrow) = \frac{1}{2\sqrt{mp^+}} 
\left[\matrix{p^+ - m \cr p^r \cr p^+ + m \cr p^r \cr}\right].
\end{equation}
They are normalized such that
\begin{equation} 
\bar{u}(p,{\lambda}) u(p,{\lambda^{\prime}}) =- 
\bar{\nu}(p,{\lambda}) \nu(p,{\lambda^{\prime}}) 
=\delta_{\lambda\lambda^{\prime}} .
\end{equation} 
The spin-$\frac{1}{2}$ projection operator is given by 
\begin{equation}
\sum_{\lambda}u(p,{\lambda} ) \bar{u}(p,{\lambda} )=\frac{(\not p +m)}{2m} .
\end{equation}
These LF spinors are related to the canonical spinors by a Melosh 
transformations.
The spin-$\frac{3}{2}$ helicity eigenstates $u^{\mu}(p,\lambda)$ are given
 in table (\ref{Table A.1}) which are normalized such that 
\be
\bar{u}^{\mu}(p,{\lambda})u^{\mu}(p,{\lambda^{\prime}})=
 -\delta_{\lambda\lambda^{\prime}}
\ee 
The spin-$\frac{3}{2}$ projection operator has the form
\be
\sum_{\lambda}u^{\mu}(p,{\lambda})\bar{u}^{\nu}(p,{\lambda})=
\frac{(\not p+m)}{2m}
\left\{ 
 -g^{\mu\nu}+\frac{2}{3}v^{\mu}v^{\nu}+\frac{1}{3}\gamma^{\mu}\gamma^{\nu} 
 +\frac{1}{3}(\gamma^{\mu}v^{\nu}-\gamma^{\nu}v^{\mu}) \right\}
\ee
\begin{table}
\begin{center}
\begin{tabular}{|c||c|c|c|c|} \hline
& & & & \\ 
$u^{\mu}(p,\lambda) $ & $u^{+}(p,\lambda) $ & $u^{-}(p,\lambda) $ & 
$u^{r}(p,\lambda) $ & $u^{l}(p,\lambda) $ 
   \\ \cline{1-1}
 $\lambda $ & & & & \\  \hline \hline
 & & & & \\
 $\frac{3}{2} $ & $ 0 $ & $ 0 $ & $ 0 $ & $ u(p,{\ua}) $ \\  
 & & & & \\ 
\hline  
 & & & & \\ 
  $\frac{1}{2} $ & $ -\frac{1}{\sqrt{3}} \frac{p^{+}}{m} u(p,{\ua}) $ & 
  $ \frac{1}{\sqrt{3}} \frac{p^{-}}{m} u(p,{\ua})$ & 
  $ 0 $ & 
  $\frac{1}{\sqrt{3}} u(p,{\da})$ \\
  & & & & \\
\hline  
 & & & & \\ 
 $-\frac{1}{2} $ & $ -\frac{1}{\sqrt{3}} \frac{p^{+}}{m} u(p,{\da}) $ &
  $ \frac{1}{\sqrt{3}} \frac{p^{-}}{m} u(p,{\da}) $ &
  $-\frac{1}{\sqrt{3}} u(p,{\ua})$ & 
  $ 0 $ \\ 
  & & & & \\
\hline  
 & & & & \\ 
 $-\frac{3}{2} $ & $ 0 $ & $ 0 $  & $ -u(p,{\da}) $ & $ 0 $ \\
& & & & \\  \hline
\end{tabular}
\caption[]{ Spin-$\frac{3}{2}$ helicity eigenstates in the Light-Front 
formalism with $u^{l}(p,\lambda)=u^{1}(p,\lambda)-iu^{2}(p,\lambda)$ and 
$u^{r}(p,\lambda)=u^{1}(p,\lambda)+iu^{2}(p,\lambda)$. } 
\label{Table A.1}
\end{center}
\end{table}
 Table (\ref{Table A.2}) contains matrix elements $\bar{u}
 (p^{\prime},{\lambda^{\prime}})
  \Gamma u(p,\lambda)$ with ($\Gamma $ = $\{ I,\gamma^+, \gamma^5$ and 
 $ \gamma^+\gamma^5 \}$). 
 In tables (\ref{Table A.3}), (\ref{Table A.4}) and (\ref{Table A.5}) 
 matrix elements 
 $\bar{u} (p^{\prime},{\lambda^{\prime}})\Gamma u^{\mu}(p,\lambda)$
  with ($\Gamma $ = $\{ I,\gamma^+ {\rm and} \gamma^5 \} $ ) respectively 
  are presented.
\begin{table}
\begin{center}
\begin{tabular}{|c||c|c|c|c|} \hline
 & & & &\\
$ \Gamma $ & $\bar{u}(p^{\prime},{\ua}) \Gamma u(p,{\ua})$ & 
$\bar{u}(p^{\prime},{\da})\Gamma u(p,{\da})$ & 
$\bar{u}(p^{\prime},{\da}) \Gamma u(p,{\ua}) $ & 
$\bar{u}(p^{\prime},{\ua}) \Gamma u(p,{\da})$ \\
 & & & &\\
   \hline \hline
 & & & & \\
 $I$ & $ {1 \over{2}} {m^{\prime} p^+ + m p^{' +} \over{p^+ p^{ ' +} }} $ &
$  {1 \over{2}} {m^{'} p^+ + m p^{+'} \over{p^+ p^{+'} }} $ &
 $ - {1 \over{2}} { p^+ p^{ ' r} -  p^{+'} p^r \over{p^+ p^{' +} }} $ &
  $  {1 \over{2}} { p^+ p^{' l } -  p^{' +} p^l \over{p^+ p^{' +} }} $ \\
  & & & & \\
$\gamma^+$ & 1 & 1 & 0 & 0 \\
 & & & & \\
 $\gamma_5$ & $ {1 \over{2}} {m^{\prime} p^+ - m p^{+'} \over{p^+ p^{+'} }} $ &
 $ - {1 \over{2}} {m^{'} p^+ - m p^{' +} \over{p^+ p^{+'} }} $ &
  $ -{1 \over{2}} { p^+ p^{ ' r} -  p^{ ' +} p^r \over{p^+ p^{' +} }} $ &
  $  -{1 \over{2}} { p^+ p^{' l } -  p^{' +} p^l \over{p^+ p^{' +} }} $ \\
  & & & & \\
  $\gamma^+\gamma^5$ & 1 &  -1 & 0 & 0 \\
  & & & & \\  \hline
  \end{tabular}
\caption[]{ The $\bar{u}(p^{\prime},{\lambda^{\prime}})
 \Gamma u(p,{\lambda})$, with $\Gamma=I$ , $\gamma^+$, $\gamma_5$ and 
 $\gamma^+\gamma_5$,  matrix elements. They are in units of
   $\sqrt{\frac{p^{+}p^{\prime +}}{m m^{\prime}}}$ . }
 \label{Table A.2}
 \end{center}
 \end{table}
\begin{table}
\begin{center}
\begin{tabular}{|c||c|c|c|c|} \hline
 & & & & \\  
  & $ \bar{u} (p^{\prime},{\ua})\Gamma u^{+}(p,\lambda)$  & 
 $ \bar{u}(p^{\prime},{\ua})\Gamma u^{-}(p,\lambda)$  &
 $ \bar{u} (p^{\prime},{\ua})\Gamma u^{r}(p,\lambda)$  &
  $ \bar{u}(p^{\prime},{\ua} )\Gamma u^{l}(p,\lambda)$ \\ 
 & & & & \\ 
 \cline{2-5} 
 & & & & \\  
 & $\bar{u} (p^{\prime},{\da})\Gamma u^{+}(p,\lambda)$  &
 $\bar{u}(p^{\prime},{\da} )\Gamma u^{-}(p,\lambda)$ &
  $ \bar{u} (p^{\prime},{\da})\Gamma u^{r}(p,\lambda)$  &
    $ \bar{u}(p^{\prime},{\da} )\Gamma u^{l}(p,\lambda)$ \\
 & & & & \\  
\hline \hline
 & & & & \\
 {$\lambda=\frac{3}{2}$} &  $0$  & $0$ & $0$ & $\pm\sqrt{2} $ \\  
  & & & & \\
  \cline{2-5}
 & & & & \\
 &  $ 0 $  & $ 0 $ & $ 0 $ & $ 0 $ \\
 & & & & \\
 \hline\hline
 & & & & \\
 {$\lambda=\frac{1}{2}$} &  $-\sqrt{\frac{2}{3}}\frac{p^{+}}{m} $ & 
    $+ \sqrt{\frac{2}{3}}\frac{p^{-}}{m} $ & 
    $0$  & $ 0 $ \\
  & & & & \\
   \cline{2-5}
   & & & & \\
  &  $ 0 $  & $ 0 $ & $ 0 $ & $\mp \sqrt{\frac{2}{3}} $ \\
  & & & & \\
   \hline\hline
& & & & \\
 {$\lambda=-\frac{1}{2}$} &  $ 0 $  & $ 0 $ &  $ \pm \sqrt{\frac{2}{3}} $ & $ 0 $ \\
& & & & \\
 \cline{2-5}
 & & & & \\
  &  $-\sqrt{\frac{2}{3}}\frac{p^{+}}{m} $ &
   $+ \sqrt{\frac{2}{3}}\frac{p^{-}}{m} $ &
  $0$  & $ 0 $ \\
   & & & & \\
\hline\hline
 & & & & \\
 {$\lambda=-\frac{3}{2}$} &  $0$  & $0$ & $0$ & $ 0 $ \\
 & & & & \\
 \cline{2-5}
 & & & & \\
&  $ 0 $  & $ 0 $ & $ \pm\sqrt{2} $ & $ 0 $ \\
 & & & & \\
  \hline
 \end{tabular}
 \caption[]{ The $\bar{u}(p^{\prime},{\lambda^{\prime}}) 
 \Gamma u^{\mu}(p,{\lambda})$ matrix elements. The lower sign is for 
 $\Gamma=\gamma^+$ and the upper sign is for $\Gamma=\gamma^+\gamma_5$. 
 They are in units of $\sqrt{\frac{p^{+}p^{\prime +}}{m m^{\prime}}}$. }
 \label{Table A.3}
\end{center}
\end{table}
\begin{table}
\begin{center}
\begin{tabular}{|c||c|c|c|c|} \hline
 & & & & \\
   & $ \bar{u}(p^{\prime},{\ua} ) u^{+}(p,\lambda)$  &
  $ \bar{u}(p^{\prime},{\ua}) u^{-}(p,\lambda)$  &
 $ \bar{u}(p^{\prime},{\ua} ) u^{r}(p,\lambda)$  &
 $ \bar{u}(p^{\prime},{\ua} ) u^{l}(p,\lambda)$ \\
& & & & \\
 \cline{2-5}
 & & & & \\
  & $\bar{u} (p^{\prime},{\da}) u^{+}(p,\lambda)$  &
  $\bar{u}(p^{\prime},{\da} ) u^{-}(p,\lambda)$ &
   $ \bar{u}(p^{\prime},{\da} ) u^{r}(p,\lambda)$  &
  $ \bar{u}(p^{\prime},{\da} ) u^{l}(p,\lambda)$ \\
  & & & & \\
  \hline \hline
   & & & & \\
   {$\lambda=\frac{3}{2}$} &  $0$  & $0$ & $0$ & $\frac{1}{\sqrt{2}}
   \frac{m^{\prime}p^{+}+m p^{\prime +}}{p^{+} p^{ ' +}} $ \\
  & & & & \\
 \cline{2-5}
 & & & & \\
  &  $ 0 $  & $ 0 $ & $ 0 $ &  $\frac{1}{\sqrt{2}}
     \frac{p^{r}p^{\prime +}-p^{\prime r}p^{+}} {p^{+} p^{ ' +}}$ \\
  & & & & \\
  \hline\hline
  & & & & \\
  {$\lambda=\frac{1}{2}$} & $-\frac{1}{\sqrt{6}}
     \frac{m^{\prime}p^{+}+m p^{\prime +}}{m p^{\prime +}} $ &
  $\frac{1}{\sqrt{6}} \frac{p^{-}}{p^{+}}
       \frac{m^{\prime}p^{+}+m p^{\prime +}}{m p^{\prime +}}$ &
   $0$  & $ \frac{1}{\sqrt{6}}
	\frac{p^{+}p^{\prime l}-p^{\prime +}p^{l}} {p^{+} p^{ ' +}}$ \\
& & & & \\
 \cline{2-5}
  & & & & \\
&  $ \frac{1}{\sqrt{6}}
    \frac{p^{+}p^{\prime r}-p^{\prime +}p^{r}} {m p^{\prime +}}$ 
    & $ \frac{1}{\sqrt{6}} \frac{p^{-}}{p^{+}}
   \frac{p^{\prime +}p^{r}-p^{+} p^{\prime r}}{m p^{\prime +}}$ & 
   $ 0 $ & $ \frac{1}{\sqrt{6}}
      \frac{m^{\prime}p^{+}+m p^{\prime +}}{p^{+} p^{ ' +}}$ \\
  & & & & \\
  \hline\hline
     & & & & \\
 {$\lambda=-\frac{1}{2}$} &  $ \frac{1}{\sqrt{6}}
     \frac{p^{l}p^{\prime +}-p^{\prime l}p^{+}} {m p^{\prime +}}$  & 
     $\frac{1}{\sqrt{6}} \frac{p^{-}}{p^{+}}
 \frac{p^{\prime l}p^{+}-p^{l} p^{\prime +}}{m p^{\prime +}}$ &
 $ -\frac{1}{\sqrt{6}}
  \frac{m^{\prime}p^{+}+m p^{\prime +}}{p^{+} p^{ ' +}}$ & $ 0 $ \\
  & & & & \\
  \cline{2-5}
& & & & \\
  &  $-\frac{1}{\sqrt{6}}
       \frac{m^{\prime}p^{+}+m p^{\prime +}}{m p^{\prime +}}$ &
  $\frac{1}{\sqrt{6}} \frac{p^{-}}{p^{+}}
	 \frac{m^{\prime}p^{+}+m p^{\prime +}}{m p^{\prime +}}$ &
   $ \frac{1}{\sqrt{6}}
	   \frac{p^{+}p^{\prime r}-p^{\prime +}p^{r}} {p^{+} p^{ ' +}}$  & 
   $ 0 $ \\
  & & & & \\
  \hline\hline
  & & & & \\
  {$\lambda=-\frac{3}{2}$} &  $0$  & $0$ & $\frac{1}{\sqrt{2}}
    \frac{p^{\prime +}p^{l}-p^{+} p^{\prime l}}{p^{+} p^{ ' +}} $ & $0$ \\
& & & & \\
 \cline{2-5}
  & & & & \\
   &  $ 0 $  & $ 0 $ &  $-\frac{1}{\sqrt{2}}
 \frac{m^{\prime }p^{+}+m p^{\prime +}} {p^{+} p^{ ' +}}$ & $ 0 $\\
   & & & & \\
  \hline
 \end{tabular}
\caption[]{ Same as table (\ref{Table A.3}) but for 
$\bar{u}(p^{\prime},{\lambda^{\prime}})
  u^{\mu}(p,\lambda)$ matrix elements. }
 \label{Table A.4}
 \end{center}
 \end{table}
\begin{table}
\begin{center}
\begin{tabular}{|c||c|c|c|c|} \hline
 & & & & \\
   & $ \bar{u} (p^{\prime},{\ua})\gamma_5 u^{+}(p,\lambda)$  &
   $ \bar{u}(p^{\prime},{\ua})\gamma_5 u^{-}(p,\lambda)$  &
    $ \bar{u}(p^{\prime},{\ua} )\gamma_5 u^{r}(p,\lambda)$  &
   $ \bar{u}(p^{\prime},{\ua} )\gamma_5 u^{l}(p,\lambda)$ \\
& & & & \\
 \cline{2-5}
  & & & & \\
  & $\bar{u}(p^{\prime},{\da} )\gamma_5 u^{+}(p,\lambda)$  &
   $\bar{u}(p^{\prime},{\da} )\gamma_5 u^{-}(p,\lambda)$ &
 $ \bar{u}(p^{\prime},{\da} )\gamma_5 u^{r}(p,\lambda)$  &
   $ \bar{u}(p^{\prime},{\da})\gamma_5 u^{l}(p,\lambda)$ \\
    & & & & \\
  \hline \hline
  & & & & \\
{$\lambda=\frac{3}{2}$} &  $0$  & $0$ & $0$ & $\frac{1}{\sqrt{2}}
  \frac{m^{\prime}p^{+}-m p^{\prime +}}{p^{+} p^{ ' +}} $ \\
  & & & & \\
   \cline{2-5}
   & & & & \\
 &  $ 0 $  & $ 0 $ & $ 0 $ &  $\frac{1}{\sqrt{2}}
  \frac{p^{r}p^{\prime +}-p^{\prime r}p^{+}} {p^{+} p^{ ' +}}$ \\
& & & & \\
  \hline\hline
   & & & & \\
  {$\lambda=\frac{1}{2}$} & $-\frac{1}{\sqrt{6}}
 \frac{m^{\prime}p^{+}-m p^{\prime +}}{m p^{\prime +}} $ &
 $\frac{1}{\sqrt{6}} \frac{p^{-}}{p^{+}}
  \frac{m^{\prime}p^{+}-m p^{\prime +}}{m p^{\prime +}}$ &
   $0$  & $-\frac{1}{\sqrt{6}}
 \frac{p^{+}p^{\prime l}-p^{\prime +}p^{l}} {p^{+} p^{ ' +}}$ \\
   & & & & \\
\cline{2-5}
  & & & & \\
  &  $ \frac{1}{\sqrt{6}}
 \frac{p^{+}p^{\prime r}-p^{\prime +}p^{r}} {m p^{\prime +}}$
 & $ \frac{1}{\sqrt{6}} \frac{p^{-}}{p^{+}}
 \frac{p^{\prime +}p^{r}-p^{+} p^{\prime r}}{m p^{\prime +}}$ &
$ 0 $ & $ -\frac{1}{\sqrt{6}}
  \frac{m^{\prime}p^{+}-m p^{\prime +}}{p^{+} p^{ ' +}}$ \\
& & & & \\
  \hline\hline
 & & & & \\
{$\lambda=-\frac{1}{2}$} &  $ -\frac{1}{\sqrt{6}}
 \frac{p^{l}p^{\prime +}-p^{\prime l}p^{+}} {m p^{\prime +}}$  &
 $-\frac{1}{\sqrt{6}} \frac{p^{-}}{p^{+}}
  \frac{p^{\prime l}p^{+}-p^{l} p^{\prime +}}{m p^{\prime +}}$ &
 $ -\frac{1}{\sqrt{6}}
  \frac{m^{\prime}p^{+}-m p^{\prime +}}{p^{+} p^{ ' +}}$ & $ 0 $ \\
& & & & \\
  \cline{2-5}
  & & & & \\
  &  $\frac{1}{\sqrt{6}}
  \frac{m^{\prime}p^{+}-m p^{\prime +}}{m p^{\prime +}}$ &
  $-\frac{1}{\sqrt{6}} \frac{p^{-}}{p^{+}}
  \frac{m^{\prime}p^{+}-m p^{\prime +}}{m p^{\prime +}}$ &
 $ \frac{1}{\sqrt{6}}
  \frac{p^{\prime +}p^{r}} {p^{+} p^{ ' +}-p^{+}p^{\prime r}}$  &
 $ 0 $ \\
 & & & & \\
   \hline\hline
  & & & & \\
   {$\lambda=-\frac{3}{2}$} &  $0$  & $0$ & $-\frac{1}{\sqrt{2}}
  \frac{p^{\prime +}p^{l}-p^{+} p^{\prime l}}{p^{+} p^{ ' +}} $ & $0$ \\
 & & & & \\
 \cline{2-5}
 & & & & \\
&  $ 0 $  & $ 0 $ &  $\frac{1}{\sqrt{2}}
 \frac{m^{\prime }p^{+}-m p^{\prime +}} {p^{+} p^{ ' +}}$ & $ 0 $\\
 & & & & \\
 \hline
 \end{tabular}
 \caption[]{  Same as table (\ref{Table A.3}) but for 
 $\bar{u}(p^{\prime},{\lambda^{\prime}})
 \gamma_5 u^{\mu}(p,\lambda)$ matrix elements.}
 \label{Table A.5}
\end{center}
\end{table}
\end{document}